\documentclass{article}
\usepackage[a4paper,margin=2cm,footskip=.5cm]{geometry}
\usepackage{setspace}
\usepackage{graphicx, epstopdf, parskip, latexsym, amsmath, amssymb, mathtools, times, amsthm, url, multicol, rotating, natbib, algorithm, algpseudocode, bm}
\usepackage[english]{babel}
\usepackage[utf8]{inputenc}
\usepackage{latexsym}
\usepackage{fancyhdr}
\usepackage{color}

\theoremstyle{plain}

\theoremstyle{definition}

\theoremstyle{remark}

\DeclareMathOperator{\Cov}{Cov}
\DeclareMathOperator{\Corr}{Corr}

\allowdisplaybreaks

\title{A note on confidence intervals for parameter estimates of\\ a spatio-temporal Ornstein-Uhlenbeck process}
\author{MICHELE NGUYEN AND ALMUT E. D. VERAART\\
	\textit{Department of Mathematics, Imperial College London}
	}
\providecommand{\keywords}[1]{\textbf{\textit{Keywords:}} #1}
\providecommand{\subjclass}[1]{\textbf{\textit{Mathematics Subject Classification:}} #1}

\date{}

\begin{document}

\maketitle
\pagenumbering{arabic}

\thispagestyle{fancy}

\begin{abstract}
We compare two ways of constructing confidence intervals for the moments-matching parameter estimates of a Gaussian spatio-temporal Ornstein-Uhlenbeck process. It was found that those obtained via pairwise likelihood approximations had lower coverages and were more prone to the curse of dimensionality as opposed to those from a parametric bootstrap procedure.
\end{abstract}

 \\

\section{Introduction}

In \cite{NV2016}, a moments-matching (MM) estimation method as well as a least-squares (LS) extension was developed for a class of spatio-temporal Ornstein-Uhlenbeck (STOU) processes. In particular, the focus was on the canonical case:
\begin{equation}
Y_{t}(x) = \int_{-\infty}^{t}\int_{x - c|t-s|}^{x + c|t-s|} \exp(-\lambda(t-s)) L(\mathrm{d}\xi, \mathrm{d}s), \label{eqn:TestOUh}
\end{equation}
where $\lambda, c > 0$ and $L$ is a homogenous L\'evy basis. The MM and LS estimators were shown to be consistent but no rate of convergence was derived. In addition, confidence intervals (CIs) of the point estimates were not considered. 
\\
Here, we consider three types of CIs for parameter estimates of a canonical STOU process:

\begin{enumerate}
\item Asymptotically normal CIs based on pairwise composite likelihood estimation;
\item Monte Carlo CIs based on diamond grid (DG) simulations from the fitted model;
\item Monte Carlo CIs based on exact simulations. 
\end{enumerate}

Although our results show that there is room for improvement, they reveal the intricacies involved in constructing CIs. The experiments presented focus on the Gaussian case where we know the exact joint distribution of our process at different space-time locations. From Example 3 in \cite{NV2016}, we know that if $L$ is Gaussian, $Y$ is Gaussian with spatio-temporal autocorrelation given by:
\begin{equation}
\Corr[Y_{t}(x), Y_{t + d_{t}}(x+d_{x})]  = \exp\left(-\lambda \max\left(|d_{t}|, \frac{|d_{x}|}{c}\right)\right). \label{eqn:STCorr}
\end{equation}
To make fair comparisons of the CI construction methods, we need to test them on exact simulations of our process. 
\\
There are various established methods to simulate from Gaussian random fields (see for example, Section 15.2 of \cite{Lantuejoul2002}). The first uses the \textit{Cholesky decomposition} of the covariance matrix: $\Sigma = MM^{T}$ where $M$ is a lower triangular $n \times n$ matrix with $n$ being the sample size. The random field $Y$ can be simulated exactly using $\bm{\mu} + M\bm{\epsilon}$ where $\bm{\mu}$ is the mean vector and $\bm{\epsilon}$ is a vector of independent standard normal random variables. Although easy to implement, this method requires intensive storage and computational time for large data sets. Based on its default memory limit, the software R can handle the case $n = 101^2$ but not $n = 151^2$. 
\\
To cope with larger data sets, many approximate methods for simulation have been suggested in the isotropic case. The turning bands method depends on the application of a central limit theorem (CLT), while classical spectral methods depend on either a CLT, an approximation of boundary effects or an approximation of the covariance function with one of compact support. A method based on circulant embedding has also been developed for the stationary and possibly non-isotropic Gaussian random fields (see for example, \cite{WC1994}). This is fast but due to the need for a non-negative definite circulant embedding of the original covariance matrix which is difficult to achieve for two-dimensional random fields, approximations are usually made. As a result, the exactness of the algorithm is lost. More recently, a remedy in the form of cut-off embedding for the isotropic case has been introduced to retain this exactness \cite[]{Gneiting2012}. This makes use of an intermediate function for distance values between $1$ and $r$, the cut-off value after which the covariance is artifically set to $0$. With differing conditions on the original covariance function, two intermediate functions and corrresponding cut-offs have been shown to lead to exactness. It is not clear what kind of extension is required for us to apply such a circular embedding strategy to Gaussian canonical STOU processes as we do not have isotropy or geometric anisotropy in the $|d_{x}^{2} + d_{t}^{2}|$ variable.
\\
Taking all of the above into account, we have chosen to conduct our tests on $101\times 101$ Cholesky data sets. In the next few sections, we go through each of the three CI construction methods and discuss the results of our simulation experiments. 

\section{Pairwise composite likelihood estimation}

\subsection{Theory for Gaussian spatial fields} \label{sec:TGRF}

Let $\{Y(\mathbf{x}), \mathbf{x}\in\mathbb{R}^{d}\}$ be a stationary Gaussian random field with mean $\mu \in\mathbb{R}$ and $\Cov(Y(\mathbf{x}), Y(\mathbf{x}')) = \sigma^{2}\rho(\mathbf{x} - \mathbf{x}'; \bm{\phi})$ where $\sigma^{2} > 0$, $\bm{\phi} \in \mathbb{R}^{p-2}$ for $p>2$ and $\rho$ is the correlation function of $Y$. Suppose that we observe the process at $n$ locations, $\mathbf{x}_{1}, \dots, \mathbf{x}_{n}$.  Following \cite{BG2015}, we can write the weighted pairwise marginal log-likelihood as a function of the parameter vector $\bm{\theta} = (\bm{\phi}, \sigma^{2}, \mu)\in \mathbb{R}^{p}$:
\begin{equation}
pl(\bm{\theta}) = \sum_{i = 1}^{n}\sum_{j>i}^{n}w_{ij} l_{ij}(\bm{\theta}), \label{eqn:wpmll}
\end{equation}
where $l_{ij}(\bm{\theta}) = -\frac{1}{2}\left[2\log \sigma^{2} + \log(1 - \rho_{ij}^{2}) + \frac{B_{ij}}{\sigma^{2}(1 - \rho_{ij}^{2})} \right]$, $\rho_{ij} = \rho(\mathbf{x}_{i} - \mathbf{x}_{j}; \bm{\phi})$, $B_{ij} = (Y(\mathbf{x}_{i}) - \mu)^{2} + (Y(\mathbf{x}_{j}) - \mu)^{2} - 2\rho_{ij}(Y(\mathbf{x}_{i}) - \mu)(Y(\mathbf{x}_{j}) - \mu)$ and $w_{ij}$ are weights used to save computational time and improve statistical efficiency. For example, we can use $w_{ij} = 1$ if $|\mathbf{x}_{i} - \mathbf{x}_{j}|\leq d$ and $0$ otherwise. 
\\
We call the vector $\hat{\bm{\theta}}$ which maximises (\ref{eqn:wpmll}) the \textit{composite likelihood (CL) estimator}. The consistency and asymptotic normality of this estimator can be established under increasing domain asymptotics and the conditions given in Appendix 1 of \cite{BG2015}. 
\\
The asymptotic variance of the CL estimator is given by $G^{-1}(\bm{\theta}) = H^{-1}(\bm{\theta})J(\bm{\theta})H^{-1}(\bm{\theta})$ where:
\begin{equation}
 H(\bm{\theta}) = - \mathbb{E}\left[\nabla^{2}pl(\bm{\theta})\right] = \sum_{i = 1}^{n}\sum_{j>i}^{n} w_{ij} \begin{pmatrix} \alpha_{ij}^{2}\kappa_{ij}\kappa_{ij}^{T} & -\frac{\rho_{ij}}{\sigma^{2}(1+\rho_{ij})}\kappa_{ij} & \mathbf{0} \\ - & \sigma^{-4} & 0\\ - & - & \frac{2}{\sigma^{2}(\rho_{ij}+1)} \end{pmatrix}, \label{eqn:Hmat}
\end{equation}
$\alpha_{ij} = (1 + \rho_{ij})^{-1}\sqrt{1 + \rho_{ij}^{2}}$, $\kappa_{ij} = (1 - \rho_{ij})^{-1}\nabla \rho_{ij}$ and $J(\bm{\theta}) = \mathbb{E}\left[\nabla pl(\bm{\theta}) \nabla pl(\bm{\theta})^{T}\right]$. 
\\
Although an explicit expression in terms of the model parameters is also available for $J(\bm{\theta})$, the computation of this is infeasible for large datasets. Thus, it is typically estimated by using a window subsampling method. Suppose that $W^{-1}J(\hat{\bm{\theta}}) \rightarrow J^{*}$ as $n\rightarrow\infty$ for some matrix $J^{*}$ and $W = \sum_{(i, j)\in S} w_{ij}$ where $S$ is the full set of observation points. Now, we have that:
\begin{equation*}
W^{-1}J(\hat{\bm{\theta}})  = \mathbb{E}\left[W^{-1} \sum_{(i, j), (i', j')\in S}w_{ij}w_{i'j'} \nabla l_{ij}(\hat{\bm{\theta}}) \nabla l_{i'j'}(\hat{\bm{\theta}})^{T}\right].
\end{equation*}
So, we can estimate $J^{*}$ by, for example, the \textit{window subsampling empirical variance (WSEV)} estimator:
\begin{equation*}
\widehat{J}^{*} = \frac{1}{m}\sum_{k = 1}^{m} \left\{ \frac{1}{W^{(k)}}\sum_{(i, j), (i', j')\in S_{k}}w_{ij}w_{i'j'} U_{ij}(\hat{\bm{\theta}}) U_{i'j'}(\hat{\bm{\theta}})^{T} \right\}. 
\end{equation*}
Here, $W^{(k)} = \sum_{(i, j)\in S_{k}} w_{ij}$, $S_{1}, \dots, S_{m}$ are the $m$ windows or subsets of the observation points determined by the subsampling method and:
\begin{equation*}
U_{ij}(\bm{\theta}) = \nabla l_{ij}(\bm{\theta}) = \begin{pmatrix} \kappa_{ij}\frac{\rho_{ij}}{1 + \rho_{ij}}\left(1 - \frac{F_{ij}}{\sigma^{2}\rho_{ij}(1 - \rho_{ij}^{2})}\right) \\ -\frac{1}{\sigma^{2}} \left(1 - \frac{B_{ij}}{2\sigma^{2}(1 - \rho_{ij}^{2})}\right) \\ \frac{2\mu}{\sigma^{2}(1 + \rho_{ij})}\left(1 - \frac{Q_{ij}}{2\mu}\right)\end{pmatrix}, 
\end{equation*}
where $F_{ij} = \rho_{ij}(Y(\mathbf{x}_{i}) - \mu)^{2} + \rho_{ij}(Y(\mathbf{x}_{j}) - \mu)^{2} - (1 + \rho_{ij}^{2})(Y(\mathbf{x}_{i}) - \mu)(Y(\mathbf{x}_{j}) - \mu)$ and $Q_{ij} = Y(\mathbf{x}_{i}) + Y(\mathbf{x}_{j})$. Conditions for the consistency of the WSEV estimator for lattice data are given in Theorem 2 of \cite{HL2000}. 
 \\
With the WSEV estimator $\widehat{J}^{*}$, the asymptotic covariance matrix of the CL estimator can be estimated by $\widehat{G^{-1}}(\hat{\bm{\theta}}) = W H^{-1}(\hat{\bm{\theta}})\widehat{J}^{*}H^{-1}(\hat{\bm{\theta}})$. The standard errors of each parameter estimate are given by the square root of the diagonal elements of $\widehat{G^{-1}}(\hat{\bm{\theta}})$. 
\\
For spatial data in $\mathbb{R}^{d}$, the optimal window size, i.e.~number of points in a window, has been shown to be $C n^{d/(d+2)}$ where $C$ is a constant and $n$ is the sample size. However, it is difficult to determine the constant $C$ for real data sets since it depends on for example, the dependence structure of the random field. In practice, computational reasons can determine the window size. In addition, using overlapping windows instead of disjoint ones can increase statistical efficiency.  
\\
In the spatial case, the consistency of the WSEV estimator typically requires that the number and size of the windows used increases simultaneously. In a spatio-temporal case where the spatial domain is fixed but the time domain increases, \cite{LGS2007} used windows in the temporal direction with full spatial coverage. Consistency of the WSEV estimator in this scenario has not yet been established. Thus, we will adapt the spatial window subsampling method for our canonical STOU process.

\subsection{Application to Gaussian canonical STOU processes}

Condition C3 for the consistency of our CL estimators in Appendix 1 of \cite{BG2015} states that the covariance function needs to have continuous second order partial derivatives with respect to our parameter vector. However, from (\ref{eqn:STCorr}), we know that when $Y$ is a canonical STOU process, its covariance function is non-differentiable with respect to $\lambda$ and $\tilde{c} = \lambda/c$. Note that we estimate $\tilde{c}$ instead of $c$ because this parameterisation seems to lead to better results in the optimisation step of the CL estimation. 
\\
In order to conduct weighted pairwise marginal likelihood estimation for our Gaussian canonical STOU process and satisfy the conditions required for asymptotic normality, we suppose instead that our correlation function is given by:
\begin{equation}
\rho\left(Y_{t}(x), Y_{t + d_{t}}(x+d_{x})\right) = \exp\left(-\lambda|d_{t}| - \tilde{c}|d_{x}|\right). \label{eqn:SSTCorr}
\end{equation}
This corresponds to a special case of the double stable model in Table 1 of \cite{PB2015}. By only using spatial and temporal pairs (that is, setting $w_{ij} = 1$ if and only if the pairs differ by space and not time and vice versa), (\ref{eqn:SSTCorr}) is equal to (\ref{eqn:STCorr}) and condition C3 is satisfied with $\bm{\theta} = (\lambda, \tilde{c}, \sigma^{2}, \mu)$. Following the notation in Section \ref{sec:TGRF}, $\sigma^{2} = \frac{c\tau^{2}}{2\lambda^{2}}$ is the variance of $Y$ and $\mu = \frac{2c\tilde{\mu}}{\lambda^{2}}$ is its mean. The parameters $\tilde{\mu}$ and $\tau^{2}$ represent the mean and variance of the L\'evy seed respectively.  Since the $w_{ij}$s are the user's choice and the $l_{ij}$s involve either the spatial or the temporal pairs (thus not differentiating between (\ref{eqn:STCorr}) and (\ref{eqn:SSTCorr})), this assumption of an alternative correlation form is valid. Now, we can find the CL estimator $\hat{\bm{\theta}}$ easily by maximising (\ref{eqn:wpmll}) using for example, the Nelder-Mead algorithm and using MM estimates as starting values. 
\\
To compute $\widehat{J}^{*}$, we treat our temporal dimension as an additional spatial dimension and use window subsampling. Note that we need to choose the window size and steps. The Delta method can be used to obtain asymptotic normal CIs for $c$, $\hat{\tau}$ and $\hat{\tilde{\mu}}$. 

\subsection{Experiment results}

Although CL estimation was shown to give good results in \cite{BSR2012} and \cite{HL2000}, in the case of the canonical STOU process, we find that it is sensitive to the dimension of the parameter space. To illustrate this, we conduct CL estimation in the following scenarios:
\begin{enumerate}
\item[A.] Estimate $\lambda$ while fixing the other parameters. 
\item[B.] Estimate $\tilde{c}$ while fixing the other parameters. 
\item[C.] Estimate $\lambda$ and $\tilde{c}$ while fixing the other parameters.  
\item[D.] Estimate $\lambda$, $\tilde{c}$ and $\sigma$ while fixing $\mu$.  
\item[F.] Estimate $\lambda$, $\tilde{c}$ and $\mu$ while fixing $\sigma$.  
\item[E.] Estimate all parameters.  
\end{enumerate}
For all the scenarios, CL estimation was applied to the same $250$ $101\times101$ Cholesky test data sets with $\lambda = c = 1$, $\tilde{\mu} = 0.2$ and $\tau = 0.1$ so that $\sigma^{2} = 0.005$ and $\mu = 0.4$. In addition, for the window subsampling, $11\times 11$ windows were used and set to move by $5$ units vertically and horizontally. Table \ref{table:CLCoverage} shows the coverage rates for the different scenarios. Comparing the results for Scenario C to those for Scenarios D-F, we see that it is hard to estimate $\sigma^{2}$ and $\mu$ while also estimating the correlation parameters. In addition, estimating $\sigma^{2}$ reduces the accuracy of the correlation parameter estimates. 
\\
A possible reason underlying the sensitivity of the CL estimation to the number of parameters to be estimated is the accuracy of the approximation of the log-likelihood. As can be seen from the conditions for the asymptotic properties, CL estimation is dependent on the mixing rate of the process $Y$ in two ways: the first is in the pairwise likelihood approximation and the second is in the estimation of $J(\bm{\theta})$. In addition, it has also been noted in \cite{BSR2012} that different window sizes do have affect the standard error estimation considerably. In their paper, the authors suggest using parametric bootstrap or Monte Carlo CIs instead. They argue that when the parameter estimators are consistent, bootstrapping will give standard error estimates of the same distribution as the empirical one. In the next section, we will use DG and Cholesky simulations to construct Monte Carlo CIs for our canonical STOU process. 

\begin{table}[htbp]
\centering
\caption{$95\%$ confidence interval coverages (in \%) based on CL estimation in the six scenarios.}
\begin{tabular}{ccccccc}
\hline
\textbf{Parameter} & \textbf{Scenario A} & \textbf{Scenario B} & \textbf{Scenario C} & \textbf{Scenario D} &  \textbf{Scenario E} & \textbf{Scenario F} \\
\hline
$\lambda$ & $76.0$ & - & $76.4$ & $14.0$ & $76.4$ & $18.8$ \\
$\tilde{c}$ & - & $80.8$ & $80.8$ & $15.5$ & $81.2$ & $17.2$ \\
$c$ & - & -  & $97.6$ & $100.0$ & $97.6$ & $100.0$ \\
$\tilde{\mu}$ & -  & - & - & - & $58.0$ & $13.2$ \\
$\tau$ & - & - & - & $57.2$ & - &  $46.8$ \\
$\mu$ & - & - & - & - & $40.4$ & $40.4$ \\
$\sigma^{2}$ & - & - & - & $50.0$ & - & $42.4$  \\
\hline
\end{tabular}
\label{table:CLCoverage}
\end{table}

\clearpage

\section{Monte Carlo CIs}

\subsection{Construction and rationale}

Suppose that we have one data set from a canonical Gaussian STOU process. Using the MM method, we can obtain point estimates for $\lambda$, $c$ as well as $\tilde{\mu}$ and $\tau$. Since we have simulation algorithms to simulate from the fitted model (that is, the STOU process with the estimated parameter values), we can generate different realisations of the fitted process. If the median of the latter estimates is close to the initial estimates, we can use quantiles of the latter estimates to get approximate CIs. 

\subsection{Using DG simulations}

In \cite{NV2016}, it was shown that DG simulations replicate the spatial correlation structure of the process better than rectangular grid simulations. Thus, we will use DG simulations to generate different realisations from the fitted model. Later, we will use Cholesky simulations for the simulations so as to reduce the effect of simulation error. However, we note that unlike the Cholesky method, DG simulation is also applicable for non-Gaussian STOU processes.
\\
Table \ref{table:MCCoverage} shows the $95\%$ CI coverages for the model parameters based on CIs constructed for $100$ Cholesky data sets using $100$ DG simulations each for the cases $\lambda = 1$ and $2$. In both cases, the other parameter settings are $c = 1$, $\tau = 0.1$ and $\tilde{\mu} = 0.2$ so that $\sigma^{2} = 0.005$ and $\mu = 0.4$. These DG simulations were generated using the same temporal grid size as the original data sets ($0.05$ units) and also contained $101\times 101$ data points. The kernel truncation parameter $p$ was set to $300$ so that the kernels were set to zero after distances of $15$ units. 
\\
From Table \ref{table:MCCoverage}, we see that except for the parameter $c$, the interval coverages for the other parameters are much lower than the expected $95\%$. Although there is a slight drop in coverage for $\mu$ and no change in coverage for $\tilde{\mu}$, the coverages improve as a whole when the true value for $\lambda$ increases from $1$ to $2$. We note that an increase in $\lambda$ is associated is a stronger mixing of the STOU process since temporal and spatial correlations decrease.

\subsection{Using Cholesky simulations}

To show that the low coverages of the DG CIs are due to the slow mixing of the process rather than simulation error, we use Cholesky decomposition to simulate from the fitted models to construct the CIs. Since the parameter $\lambda$ affects both the temporal and the spatial correlation of our canonical STOU process, we compare the $95\%$ confidence interval coverages for the cases $\lambda = 1$, $2$ and $4$. A larger $\lambda$ value implies weaker correlation and stronger mixing. This should lead to faster convergence to the true parameter values and thus higher coverage rates. Indeed, we find that in Table \ref{table:MCCoverage}, the $95\%$ CI coverages improve massively when $\lambda$ is increased from $1$ to $4$.
\\
Although using DG and Cholesky simulations give similar coverage results for the case $\lambda = 2$, when we have Gaussian STOU processes, it is worth using the Cholesky method not only because it gives exact simulations but also because the computation of a Monte Carlo CI takes around seven minutes as opposed to the one and a half hour required by the DG method on a PC with characteristics: Intel$^{\circledR}$ Core\texttrademark i7-3770 CPU Processor @ 3.40GHz; 8GB of RAM; Windows 8.1 64-bit .   

\clearpage

\begin{table}[tbp]
\centering
\caption{$95\%$ confidence interval coverages (in \%) when we use DG and Cholesky simulations. The coverages are computed based on $100$ data sets for each case. In all the cases, the other parameters are set to $c = 1$, $\tau = 0.1$ and $\tilde{\mu} = 0.2$ so that $\sigma^{2} = 0.005$ and $\mu = 0.4$.}
\begin{tabular}{cccccc}
\hline
\textbf{Parameter} & \textbf{DG} ($\lambda = 1$) & \textbf{DG} ($\lambda = 2$) & \textbf{Cholesky} ($\lambda = 1$) & \textbf{Cholesky} ($\lambda = 2$) & \textbf{Cholesky} ($\lambda = 4$) \\
\hline
$\lambda$ & $57$ & $81$ & $63$ & $78$ & $91$ \\
$c$ & $100$ & $100$ & $94$ & $96$ & $94$ \\
$\tilde{\mu}$ & $74$ & $74$ & $62$ & $78$ & $90$ \\
$\tau$ & $66$ & $85$ & $62$ & $78$ & $84$ \\
$\mu$ & $74$ & $68$ & $86$ & $91$ & $89$ \\
$\sigma^{2}$ & $76$ & $100$ & $65$ & $78$ & $87$ \\
\hline
\end{tabular}
\label{table:MCCoverage}
\end{table}

\begin{figure}[tbp]
\centering
\caption{Schematic for coverage proxy terms.}
\label{fig:coverageproxy}
\includegraphics[width = 3.6in, height = 1.5in, trim = 0.4in 0.2in 0.4in 0in]{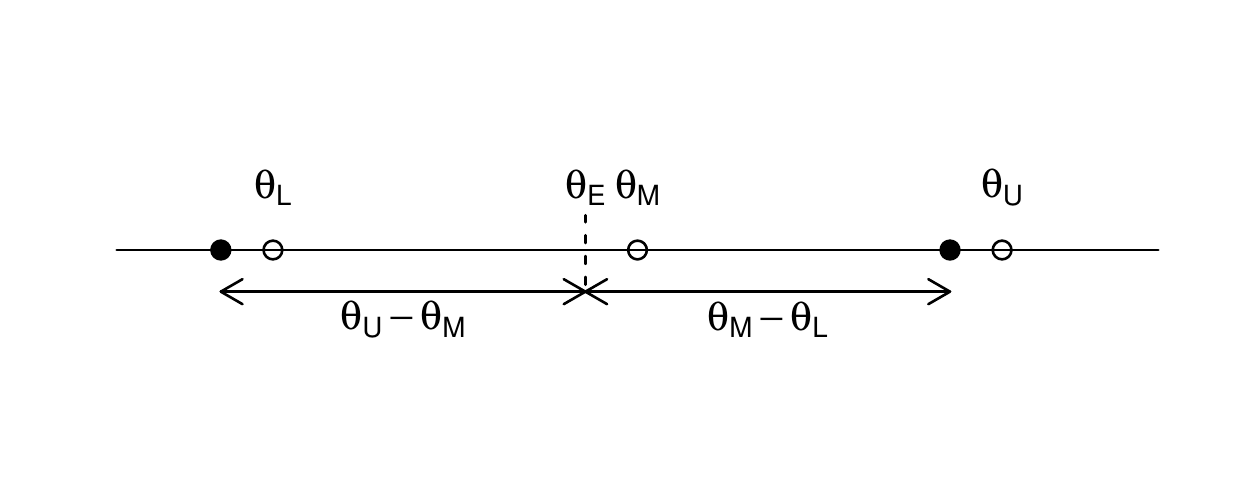}
\end{figure}

\begin{table}[tbp]
\centering
\caption{$95\%$ confidence interval coverage proxies (in \%) when we use Cholesky simulations for $\lambda = 1, 2, 4$. The proxies are computed based on the empirical cumulative distribution of the MM estimates from $100$ data sets. In all the cases, the other parameters are $c = 1$, $\tau = 0.1$ and $\tilde{\mu} = 0.2$ so that $\sigma^{2} = 0.005$ and $\mu = 0.4$.}
\begin{tabular}{cccc}
\hline
\textbf{Parameter} & $\lambda = 1$ & $\lambda = 2$ & $\lambda = 4$ \\
\hline
$\lambda$ & $79.6$ & $85.1$ & $92.0$ \\
$c$ & $100.0$ & $100.0$ & $100.0$ \\
$\tilde{\mu}$ & $98.4$ & $100.0$ & $100.0$ \\
$\tau$ & $100.0$ & $100.0$ & $100.0$ \\
$\mu$ & $100.0$ & $100.0$ & $100.0$ \\
$\sigma^{2}$ & $100.0$ & $100.0$ & $100.0$ \\
\hline
\end{tabular}
\label{table:CoverageProxy}
\end{table}

\clearpage

\subsection{Some practical guidance}

\begin{figure}[tbp]
\centering
\caption{Boxplots of the MM estimates for $100$ Cholesky data sets. The results for the case $\lambda = 1, 2, 4$ are shown in first, second and third columns respectively. The true parameter values are denoted by the red horizontal lines.}
\label{fig:bpCL}
\includegraphics[width = 3.2in, height = 9in, trim = 0.4in 0.2in 0.4in 0in]{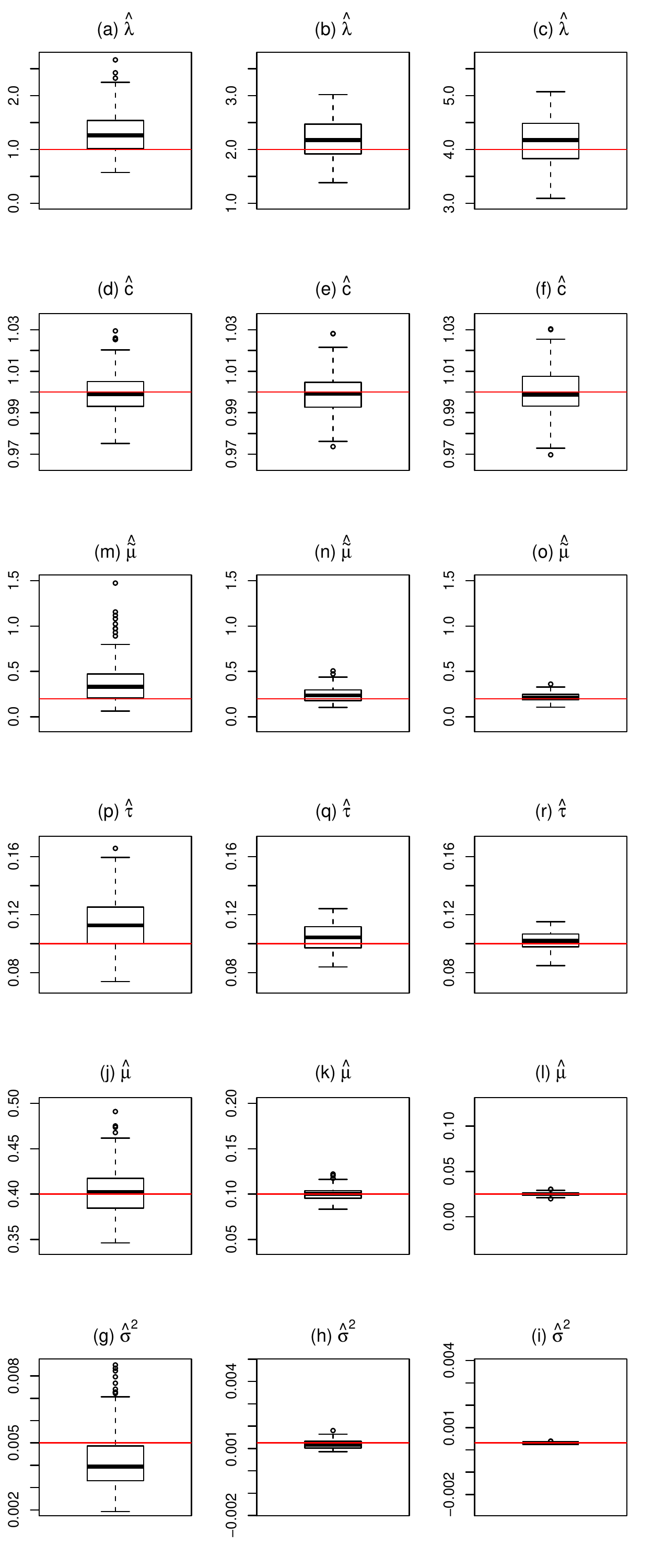}
\end{figure}

From our analysis, we see that the value of $\lambda$ affects the coverage rates of the Monte Carlo CIs. This is likely due to the rate at which convergence of the MM estimates to their true values occur. A higher value of $\lambda$, relative to the other parameter values, leads to stronger mixing of the process and faster convergence. This lends better justification for Monte Carlo CIs. While one can rescale the temporal and spatial units of our data to some degree in order to achieve higher $\lambda$ estimates, this leads to a reduction in the area covered by our data and is in conflict with the increasing domain asymptotics underlying some theoretical properties of our estimators. 
\\
One possible way to check if good coverage can be obtained is to calculate coverage proxies based on the empirical distribution of the MM estimates from the fitted model. For example, if we want to compute a proxy for the $95\%$ confidence interval coverage of the Monte Carlo CI for $\theta$, we can define $\theta_{L}$, $\theta_{M}$ and $\theta_{U}$ to be the $2.5^{\text{th}}$ quantile, median and $97.5^{\text{th}}$ quantile of its MM estimates from fitted model simulations. The initial MM estimate used to fit the model, that is the true parameter value of the fitted model, is denoted by $\theta_{E}$. A schematic for these terms is given in Figure \ref{fig:coverageproxy}. 
\\
We work under the assumption that further simulations using a MM estimate along the range represented by the horizontal line would lead to estimates which have a similar distribution and relation to the true value as the one outlined by $\theta_{L}$, $\theta_{M}$, $\theta_{U}$ and $\theta_{E}$. In Figure \ref{fig:coverageproxy}, the filled black dots represent the leftmost and rightmost points for which $\theta_{E}$ lies within a 95\% CI. Thus, they represent the extreme values that contribute to the calculation of the coverage. With this in mind, a simple coverage proxy can be calculated via:
\begin{equation*}
CP = ECDF(\theta_{E} + (\theta_{M} - \theta_{L})) - ECDF(\theta_{E} - (\theta_{U} - \theta_{M})), 
\end{equation*}
where $ECDF$ is the empirical cumulative distribution function of the MM estimates. 
\\
Table \ref{table:CoverageProxy} shows the results when we apply this proxy to Cholesky simulations for $\lambda = 1, 2, 4$. Since the method relies on the empirical distribution of the MM estimates, we also show the box plots of these in Figure \ref{fig:bpCL}. From Table \ref{table:CoverageProxy}, we see that the coverage proxies tend to overestimate the coverages calculated by the experiments in Table \ref{table:MCCoverage}. This is likely to be because we have not fully accounted for randomness in the further simulations as well as the differing accuracy of the MM estimation for different parameter values. With reference to Figure \ref{fig:bpCL}, the extent of this overestimation seems to increase as the range of the estimates narrows since $\tau$ and $\sigma^{2}$ suffer more from this than the other parameters. Despite these limitations, we see that this simple coverage proxy can distinguish between the better coverage scenarios because the proxy for $\lambda$ increases steadily as $\lambda$ increases from $1$ to $4$.
\\
As an alternative to check the coverage properties of Monte Carlo CIs in practice, we can conduct coverage experiments with the fitted model acting as the true one. This, however, will be more computationally expensive.

\clearpage

\section{Conclusion and Discussion}

In this note, we have investigated three different ways of constructing confidence intervals (CIs) for the parameters of a canonical spatio-temporal Ornstein-Uhlenbeck (STOU) process. Focusing on Gaussian processes, we showed that the asymptotically normal CIs obtained from pairwise composite likelihood (CL) estimation has poor coverage properties especially when the number of parameters to be estimated increases. Furthermore, due to the optimisation step in the CL estimation, the performance is also affected by the way we parameterised the model. 
\\
Better coverage results were obtained when we used parametric bootstrap or Monte Carlo CIs involving diamond grid (DG) or Cholesky simulations and moments-matching (MM) estimation. While the Cholesky method is only applicable for Gaussian STOU processes, the DG algorithm can be used for non-Gaussian cases. We showed that coverage rates for these Monte Carlo CIs are dependent on the mixing rate of the process which in turn affects the consistency of the MM estimators. When using the DG simulations in practice, one should also check if the choice of the grid size and kernel truncation parameter is appropriate for the estimated correlation structure so as to reduce the simulation error associated with the DG algorithm. 
\\
For an empirical example, one can imitate coverage experiments by simulating from the fitted model. However, this is computationally expensive and there is ground for developing more sophisticated coverage proxies than the one presented in Section 3.4. Another area for future research would be to consider the use of the rectangular grid algorithm and the least-squares adaptation of the MM method to construct Monte Carlo CIs for more general STOU processes with non-linear integration sets. 

\bibliographystyle{agsm} 
\bibliography{refs}

\end{document}